\documentstyle[epsfig,graphics]{aipproc}
\input pstricks
\input ulem.sty
\input times.sty
\begin{document}
\def\A{\mbox{ASCA\ }}   \def\B{\mbox{BeppoSAX\ }}   \def\ind{\mbox{~~~}}
\def\asm{\mbox{RXTE/ASM\ }} \def\exo{\mbox{EXOSAT\ }}
\def\etal{{et al. }}
\newcommand{\lx}{\mbox{L$_X$}} \newcommand{\ergs}{\mbox{erg s$^{-1}$}}
\newcommand{\ergcm}{\mbox{erg cm$^{-2}$ s$^{-1}$}}
\newcommand{\mrka}{\mbox{$^{1}$}}
\newcommand{\mrkb}{\mbox{$^{2}$}}
\newcommand{\mrkc}{\mbox{$^{3}$}}
\newcommand{\mrkd}{\mbox{$^{4}$}}
\newcommand{\mrke}{\mbox{$^{5}$}}
\newcommand{\mrkf}{\mbox{$^{6}$}}

\title{
Cyclotron lines in X-ray pulsars as a probe of
relativistic plasmas in superstrong magnetic fields}

\bigskip\medskip

\author{
D. Dal Fiume\mrka, Filippo Frontera\mrkb,
Nicola Masetti\mrka, Mauro Orlandini\mrka, Eliana Palazzi\mrka,
Stefano Del Sordo\mrkc, Andrea Santangelo\mrkc, Alberto Segreto\mrkc,
Tim Oosterbroek\mrkd, Arvind N. Parmar\mrkd}

\address
{ \mrka Istituto TESRE/CNR, via Gobetti 101, 40129 Bologna, Italy \\
\mrkb Istituto TeSRE and Dipartimento di Fisica, Universit\'a di
Ferrara, via Paradiso 1, 44100 Ferrara, Italy\\
\mrkc IFCAI/CNR, via U. La Malfa 153, 90146 Palermo, Italy\\
\mrkd Space Science Department, ESA, ESTEC, Noordwjik, The
Netherlands
}

\maketitle
\begin{abstract}
The systematic search for the presence of cyclotron lines in the spectra
of accreting X-ray pulsars is being carried on with the BeppoSAX
satellite since the beginning of the mission. These highly successful
observations allowed the detection of cyclotron lines in many of
the accreting X-ray pulsars observed. Some correlations between
the different measured parameters were found. We present these
correlations and discuss them in the framework of the current
theoretical scenario for the X--ray emission from these sources.
\end{abstract}

\section*{Introduction}

Accreting magnetized neutron stars are an ideal cosmic
laboratory for high energy relativistic physics. Cyclotron resonance
features are the signature of the presence of a superstrong magnetic field,
following the first discovery in Her X-1 (Tr\"umper et al.
\cite{trump}). These features are due to the discrete Landau energy levels
for motion of free electrons perpendicular to the field in presence
of a locally uniform superstrong magnetic field. A slight deviation
from a pure harmonic relationship in the spacing of the different levels
is expected due to relativistic effects
($\frac{\omega_n}{m_e} =((1+2n\frac{B}{B_{\rm crit}}
\sin^2\theta)^\frac{1}{2} -1)/\sin^2 \theta$, e.g. Araya and Harding 
\cite{araya}).
Therefore the detection of these features in the emitted X--ray spectra
is in principle a direct measure of the field intensity.\\
As the number of sensitive measurements in the hard X--ray interval (above
$\sim$ 10 keV) is continuously growing, a sample is available to search
for possible correlations between the observed parameters.
A detailed modeling is difficult and a parametrized shape of the
continuum still is not available from theoretical models, but
substantial advances in our understanding of the radiation transport in
strongly magnetized atmospheres were done in the last decade (e.g.
Alexander et al. \cite{alex1}, Alexander and M\'esz\'aros \cite{alex2},
Araya and Harding \cite{araya}, Isenberg et al. \cite{isenba,isenbb},
Nelson et al. \cite{nelson}).
Some of these new results focused on the properties of the cyclotron
resonance features observed in the spectra of accreting X--ray pulsars. 
In this report we discuss the current status of the measurements of
cyclotron lines, with emphasis on the possible correlations between
observable parameters.

\section*{The data}

The \B satellite has observed all the bright persistent
and three bright transient (recurrent) accreting X--ray pulsars.
Apart from the case of X Persei (Di Salvo et al. \cite{robba}), a source
with a luminosity substantially lower
than the other sources in the sample, the spectra observed by \B can be
empirically described using the classical power--law--plus--cutoff spectral
function by White et al. \cite{white}. The sensitive broad band \B
observations also allowed the detailed characterization of low energy
components below a few keV
(like in Her X--1, Dal Fiume et al. \cite{herx1}, and in 4U1626--67,
Orlandini et al. \cite{1626}) and the detection of absorption
features in the hard X--ray range of the spectra, interpreted as cyclotron
resonance features.

A summary of the properties of the broad band spectra and of the
cyclotron lines as measured with \B is reported in Dal Fiume et al.
\cite{ddf}.
From these measurements we obtained evidence of a correlation between
the centroid energy of the feature and its width. This correlation is
presented and discussed elsewhere (Dal Fiume et al. \cite{ddf,ddf98}).

\subsection*{Transparency in the line}
A straightforward parameter to be obtained from observations is the
transparency in the line, defined as the ratio between the transmitted
observed flux and the integrated flux from the continuum without the
absorption feature. This ratio likely depends on the harmonic
number of the feature we are observing (e.g. Wang et al. \cite{wang})
and on the physical
parameters of the specific accretion column. From an observational
point of view, this ratio is strongly affected by the modelization of
the ``continuum'' shape, that is by the spectral shape used to
describe the differential broad band photon number spectrum. In Figure 1
we report the observed transparencies obtained dividing the observed
by the expected flux, both integrated in a $\pm 2\sigma$ interval around
the line
centroid (here $\sigma$ is the Gaussian width of the measured cyclotron
feature). To further emphasize the uncertainty in this estimate, we
added a 10\% error to the data. The purely statistical uncertainties are
substantially smaller.
\begin{figure}[h]
\centerline{
\psfig{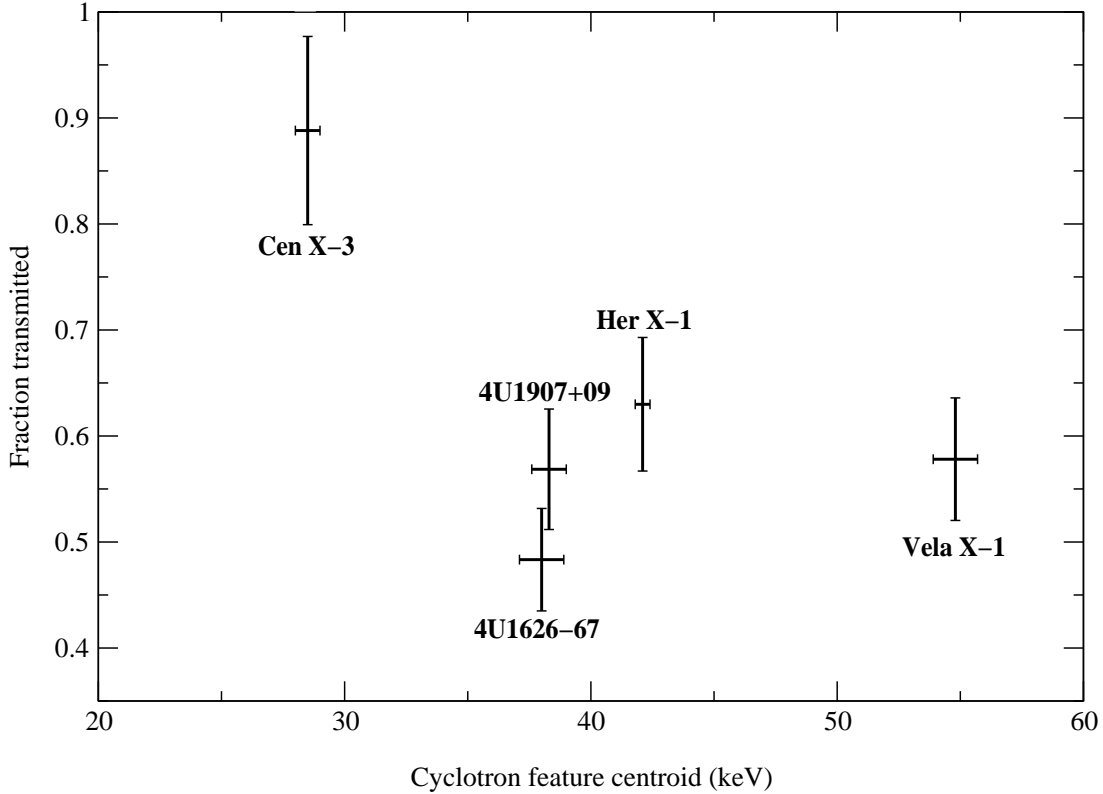}}
\caption{Transparency in the observed cyclotron resonance features
with \B. The error bars are {\it NOT} statistical, but rather indicate
the uncertainty in the determination of the shape and intensity of the
expected continuum flux (with no line absorption).}
\end{figure}
\begin{figure}[h]
\centerline{
\psfig{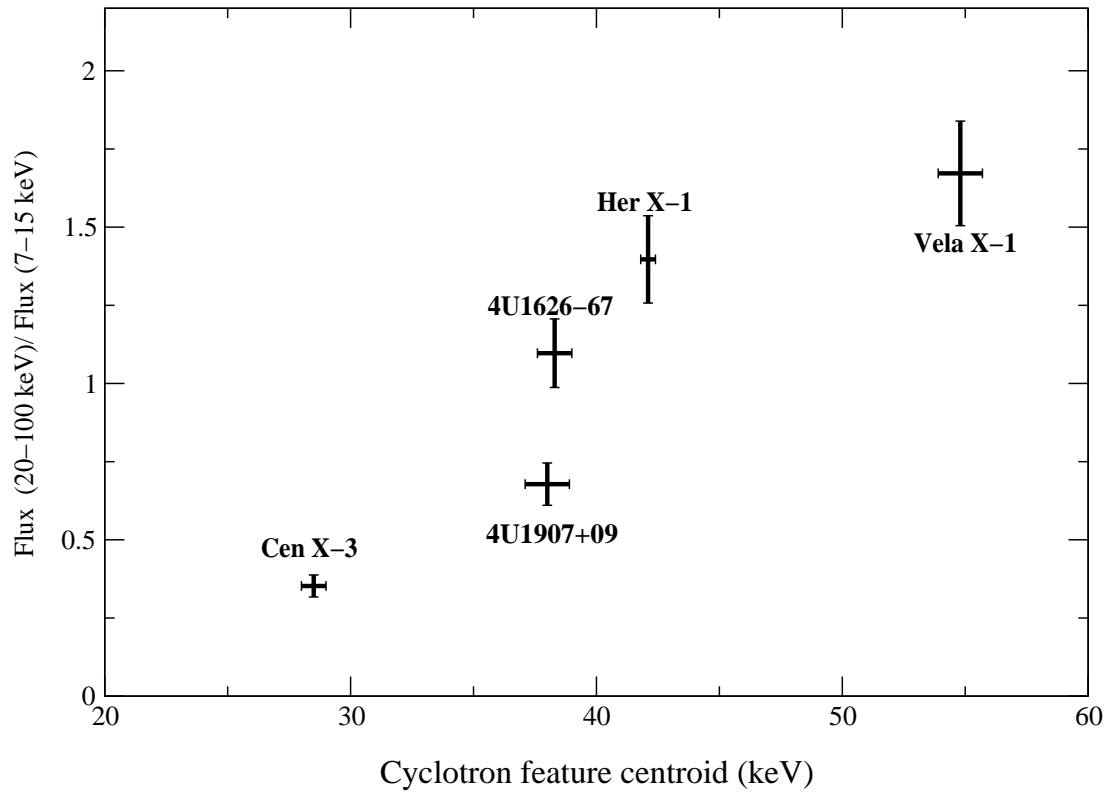}}
\caption{Ratio between the measured photon flux in two energy band
versus the cyclotron line energy
with \B. The error bars are {\it NOT} statistical, but rather indicate
the uncertainty in the determination of the shape and intensity of the
measured photon flux.}
\end{figure}
This measured transparency is related to a simple physical parameter,
the opacity to photons with energy approximately equal to the cyclotron
resonance energy. However the emerging integral flux and the shape of
the line itself are non trivially related to the radiation transport in
this energy interval, a rather difficult problem to be solved.\\
From the phenomenological point of view, one can observe that the
measured transparencies cluster around 0.5--0.6, with the notable
exception of Cen X--3.

\subsection*{Magnetic field intensity and spectral hardness}
The influence of the magnetic field intensity on the broad band spectral
shape is debated. Early attempts to estimate a possible dependence of
electron temperature, and therefore of broad band spectral shape, on the
magnetic field intensity were done by Harding et al \cite{hard84}.
Actually they conclude that {\it ``the equilibrium atmospheres have
temperatures and optical depths that are very sensitive to the strength
of the surface magnetic field''}. If this is the case and if the broad
band spectral hardness is related, as one could naively assume, to
the temperature of the atmosphere, some correlation between this
hardness and the cyclotron line energy should appear in data.

This seems to be the case shown in Figure 2. Here we report the ratio
between photon fluxes in two ``hard'' bands (the flux in 20--100 keV
divided by the flux in 7--15 keV) versus the cyclotron line centroid.
The ratio between the two fluxes is affected by the
choice of the continuum, as in Figure 1. We therefore also in this case
added 10\% error bars that indicate this uncertainty. The
statistical errors are substantially smaller.\\
The number of sources in this plot is still very limited and therefore
one cannot
exclude that this apparent correlation is merely due to the limited size
of the sample. Nevertheless the apparent correlation is in the right
direction, i.e. harder spectra are observed for higher field
intensities.\\
We parenthetically add that no cyclotron resonance feature was observed
in the pulse--phase averaged spectra of the two hardest sources of this
class observed
with \B (GX1+4 Israel et al. \cite{israel} and GS1843+00 Piraino et al.
\cite{piraino}). If this correlation proves to be correct, this may
suggest that cyclotron resonance features in these two sources should be
searched at the upper limit of the \B energy band or beyond.

\subsection*{Conclusions}
In conclusion, even if no complete, parametric theoretical approach to
model the observed spectra of accreting X--ray pulsars is still
available, some quantitative measures of parameters of hot plasmas in
superstrong magnetic fields are possible.\\
Modeling the transparency in the cyclotron resonance feature is a
complex problem. Further information will be extracted
from maps of this transparency as a function of pulse phase. \\
The correlation between
spectral hardness and field intensity is in agreement
with theoretical models. This correlation, if confirmed, can be used as
a rough estimate of the magnetic field intensity from the measured
spectral hardness.

\acknowledgments
This research is supported by Agenzia Spaziale Italiana (ASI) and 
Consiglio Nazionale delle Ricerche (CNR) of Italy. \B is a joint
program of ASI and of the Netherlands Agency for Aerospace Programs (NIVR).

\end{document}